\documentclass[runningheads]{llncs}
\usepackage[T1]{fontenc}
\usepackage{graphicx}
\begin{document}
\title{Enhancing Eye-Tracking Performance through Multi-Task Learning Transformer}
\titlerunning{Enhanced Eye-Tracking via Multi-Task Learning Transformers}
%
\author{Weigeng Li \and
Neng Zhou \and
Xiaodong Qu}
%
%

\institute{The George Washington University, Washington, DC 20052, US \\
\email{\{weigengli, nengzhou, x.qu\}@gwu.edu}}

\maketitle              
\begin{abstract}
In this study, we introduce an innovative EEG signal reconstruction sub-module designed to enhance the performance of deep learning models on EEG eye-tracking tasks. This sub-module can integrate with all Encoder-Classifier-based deep learning models and achieve end-to-end training within a multi-task learning framework. Additionally, as the module operates under unsupervised learning, it is versatile and applicable to various tasks. We demonstrate its effectiveness by incorporating it into advanced deep-learning models, including Transformers and pre-trained Transformers. Our results indicate a significant enhancement in feature representation capabilities, evidenced by a Root Mean Squared Error (RMSE) of 54.1mm. This represents a notable improvement over existing methods, showcasing the sub-module's potential in refining EEG-based model performance.

The success of this approach suggests that this reconstruction sub-module is capable of enhancing the feature extraction ability of the encoder.  Due to the sub-module being mounted as a sub-task under the main task and maintained through a multi-task learning framework, our model preserves the end-to-end training process of the original model. In contrast to pre-training methods like autoencoder, our model saves computational costs associated with pre-training and exhibits greater flexibility in adapting to various model structures. Benefiting from the unsupervised nature of the sub-module, it can be applied across diverse tasks.
We believe it represents a novel paradigm for improving the performance of deep learning models in EEG-related challenges.

\keywords{EEG Eye-Tracking
            \and Hybrid Vision Transformers 
            \and Multi-Task Learning 
            \and Signal Reconstruction 
            \and Unsupervised Learning
            \and Spatio-Temporal Data Processing
            \and Feature Extraction
            \and Neuroscience }
\end{abstract}

\section{Introduction}

Electroencephalography (EEG) stands as a crucial neuroimaging tool for comprehending the complex workings of brain activity and neural interactions \cite{teplan2002fundamentals}. EEG captures the electrical signals produced by neurons, providing a distinctive view of the brain's dynamic processes with exceptional temporal precision. A wide range of machine learning and deep learning techniques have been applied to EEG data, facilitating advanced understanding and applications across multiple domains \cite{craik2019deep,roy2019deep,altaheri2023deep,gao2021complex,hossain2023status,key2024advancing,li2024enhancing,koome2023trends,murungi2023empowering,qu2022eeg4home,yi2022attention,dou2022time,zhou2022brainactivity1,wang2022eeg,qu2022time,qu2020identifying,qu2020using,qu2020multi,qu2018eeg,qu2019personalized}.

EEG has been employed in various tasks, reflecting its versatility. Researchers have harnessed EEG data for purposes such as brain-computer interfaces (BCIs) \cite{ang2013brain}, sleep analysis \cite{motamedi2014signal}, and more recently, eye movement prediction \cite{kastrati2021eegeyenet}. These tasks have revealed different aspects of brain function and have contributed to our understanding of the neural mechanisms underlying cognition, behavior, and sensory processing.

The challenges of EEG-based tasks are multiple, including issues related to data quality, computational complexity, and model generalization \cite{rashid2020current}. EEG signals are vulnerable to noise and artifacts, which can affect the reliability of results. In addition, the high dimension of EEG data poses computational challenges, requiring sophisticated preprocessing and feature extraction techniques.

In recent years, convolutional neural networks (CNNs) have become a powerful tool in the field of EEG research \cite{craik2019deep}. Originally developed for image analysis, convolutional neural networks have now been applied to process and interpret EEG data. These neural networks can automatically learn complex spatial and temporal patterns in EEG signals, providing a new dimension in the analysis of brain activity. 

Multi-task Learning (MTL) \cite{caruana1997multitask}, in contrast to single-task learning (STL), involves simultaneous consideration of multiple related tasks, leveraging shared information to address complex challenges. This approach capitalizes on task connections to extract complementary information, enhancing decoding model accuracy and reliability. Previous research has highlighted the advantages of multi-task EEG analysis, revealing its applications in emotion recognition \cite{li2022emotion} \cite{choo2023effectiveness}, classification \cite{autthasan2021min2net} \cite{song2019eeg}, and disease prediction \cite{ma2018predicting}.  

\subsection{Research Questions}

Decoding EEG signals typically involves a series of steps, including preprocessing, feature extraction, and classification. Achieving successful EEG decoding in open-world scenarios necessitates careful consideration at each stage. Even when recorded under the most stringent conditions  \cite{chen2022toward}, EEG signals are susceptible to various artifacts such as eye blinks, muscle interference, cardiac disturbances, and electromagnetic interference. 

In this context, MTL emerges as a valuable strategy for improving the feature-extracting ability of EEG decoding. By harnessing the power of multiple related tasks, MTL enhances the generalization capabilities of EEG models and mitigates the risk of overfitting, thereby contributing to more effective EEG signal analysis in open-world environments. This approach leverages the inherent connections between tasks and allows for the extraction of complementary information, ultimately enhancing the accuracy and reliability of EEG decoding models.

Our research aims to address the following questions at the intersection of Machine Learning and EEG eye-tracking:
\begin{enumerate}
    \item Can we use EEG Signal Reconstruction as a sub-task to enhance the Transformer encoder's feature-extracting ability?
           
    \item Which aspects of the prediction results, like specific regional accuracy or the overall prediction pattern, improved after integrating our framework?
    
\end{enumerate}

\section{Related work}

\subsection{Deep Learning for EEG Tasks}
Early studies highlighted the potential of Convolutional Neural Networks (CNNs) in EEG analysis. For instance, a CNN-based approach \cite{mao2020eeg} is introduced for epileptic seizure classification on EEG data, utilizing the continuous wavelet transform (CWT) to convert EEG data into time-frequency domain images. Similarly, Transformer-based models \cite{sun2021eeg} have shown their superiority over CNNs, RNNs, and DBNs in EEG classification, indicating the promise of the hybrid Transformer-CNN approach. 

\subsection{MTL for EEG Tasks}

Multi-task Learning (MTL) \cite{caruana1997multitask} has been leveraged in various EEG signal analysis applications, including emotion recognition \cite{li2022emotion} \cite{choo2023effectiveness}, classification \cite{autthasan2021min2net} \cite{song2019eeg}, and disease prediction \cite{ma2018predicting}. DMTL-BCI \cite{song2019eeg} employed an MTL framework to jointly optimize three modules (representation, classification, and reconstruction), outperforming state-of-the-art methods by 3.0\% on BCI Competition IV dataset 2a. MIN2Net \cite{autthasan2021min2net} utilized deep metric learning and autoencoder for subject-independent motor imagery EEG signal classification, outperforming state-of-the-art techniques by 6.72\% and 2.23\% on the SMR-BCI and OpenBMI datasets. Choo et al. \cite{choo2023effectiveness} investigated the effectiveness of MTL in raw EEG-based emotion recognition, demonstrating significant classification accuracy improvements with their MTL-ShallowConvNet architecture. Furthermore, EEG-DEMTL \cite{cheng2023evolutionary} is a computation-based MTL network for assessing railway passenger comfort through EEG signals, improving the evaluation performance by 6.3\% in field experiments.

\subsection{Vision Transformers (ViTs)}

\begin{figure}[t]
    \centering
    \includegraphics[width=\textwidth]{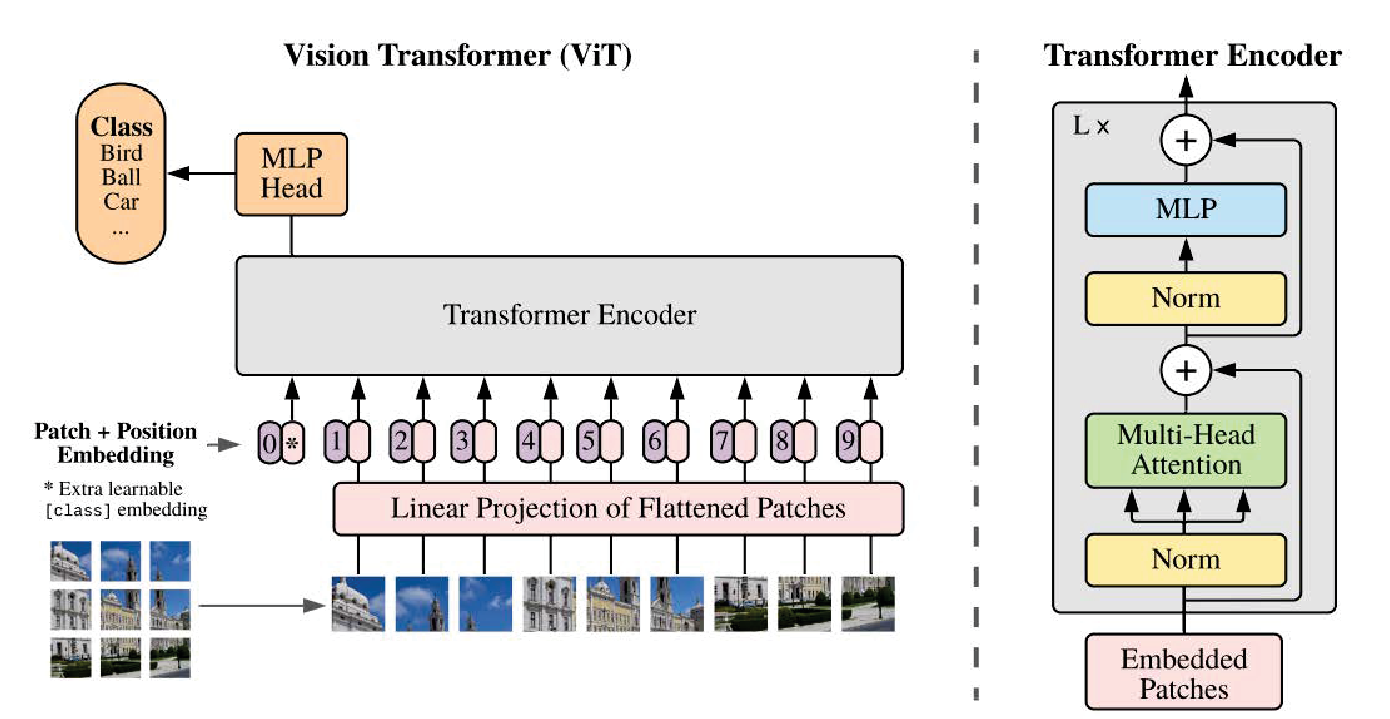}
    \caption{Vision Transformer Encoder proposed by \cite{dosovitskiy2020image}}
\label{fig05}
\end{figure}

The transformer model \cite{vaswani2017attention} is a deep learning model based on the self-attention mechanism, primarily used for processing sequential data. The core idea of the Transformer model is that through the attention mechanism, the model can focus on any part of the input sequence, thereby more effectively capturing long-distance dependencies within the sequence. This mechanism has led to tremendous success for the Transformer in the field of natural language processing (NLP), especially in tasks such as machine translation, text generation, and comprehension.

Building on the success of the Transformer model, Dosovitskiy and others proposed the Vision Transformer (ViT) in 2020 \cite{dosovitskiy2020image} Vision Transformer. ViT applies the concept of the Transformer to the field of computer vision, dividing images into a series of small patches and feeding these patches as a sequence into a self-attention-based Transformer network. This approach allows ViT to process image data effectively, capturing complex patterns and relationships within images, thus achieving excellent performance in image classification and other visual tasks. Subsequently, ViT has also shown great potential in other areas, such as EEG data analysis, demonstrating its effectiveness in processing non-traditional visual data.

Several studies have demonstrated their effectiveness regarding the application of Vision Transformers (ViT) in EEG tasks.  Yang and Modesitt demonstrated the application of a hybrid ViT model, pre-trained on ImageNet, in an EEG regression task. Additionally, a bi-branch Vision Transformer-based EEG emotion recognition model, Bi-ViTNet, integrating spatial-temporal and spatial-frequency feature representations, has shown ViT's potential in handling complex EEG data\cite{10098561}.
EEG-ConvTransformer demonstrated improved classification accuracy over state-of-the-art techniques in five different visual stimuli classification tasks. This further proves the effectiveness of ViT models in EEG signal processing.\cite{bagchi2021eegconvtransformer} Finally, the importance of the attention mechanism in EEG signals was introduced through two ViT-based methods for the classification of EEG signals based on emotions.\cite{9629837}

These studies indicate that ViT models can effectively process EEG data, especially in complex tasks such as emotion recognition and visual stimuli classification. These findings support the use of ViT as a base model for multi-task learning (MTL) in EEG tasks.

\section{Methods}

In this paper, we plan to combine multi-task learning and Vision Transformer \cite{dosovitskiy2020transformers} to enhance the performance of the EEGEyeNet dataset's eye-tracking task \cite{kastrati2021eegeyenet}. By simultaneously addressing multiple related tasks within the dataset, we aim to improve the model's performance on the eye-tracking task. Our approach holds the potential to uncover novel connections and enhance the overall understanding of eye-tracking patterns in the context of EEG signals.

\subsection{Model Architecture}

 \begin{figure}[t]
    \centering
    \includegraphics[width=\textwidth]{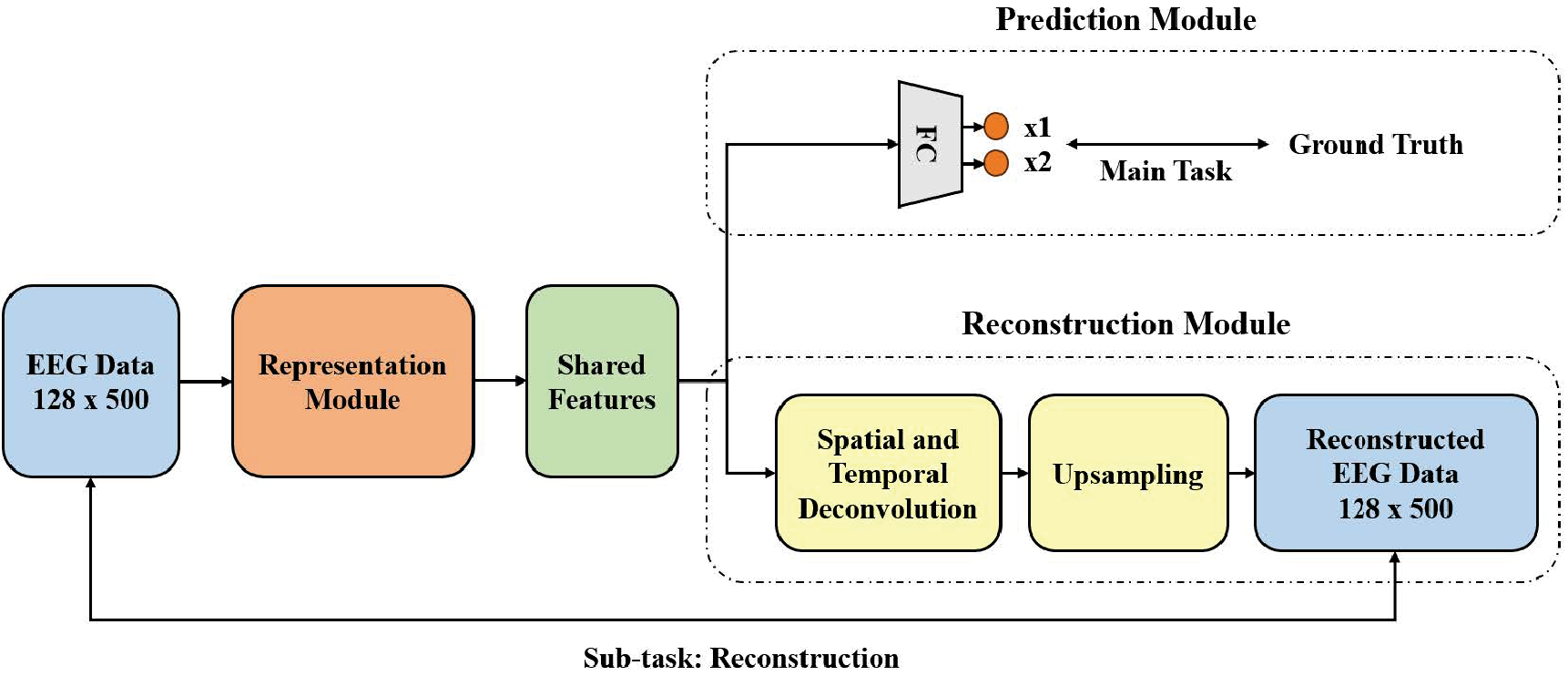}
    \caption{Proposed MTL-Transformer Architecture: Eye Tracking and Data Reconstruction }
\label{fig02}
\end{figure}

Our model architecture is specifically designed to enhance performance in EEG eye-tracking tasks. The cornerstone of our approach is the introduction of a multi-task framework, which handles various sub-tasks simultaneously. This design choice is motivated by the need to capture the diverse aspects of EEG data more effectively.

Drawing inspiration from the work of Song et al. \cite{song2019eeg}, our Multi-task Learning Transformer uniquely combines classification and reconstruction tasks within its architecture. By doing so, it efficiently leverages the representation module to maintain dual capabilities in feature extraction. This multi-task learning approach significantly boosts the model's ability to generalize across different EEG data scenarios.

The processing flow of our model, particularly highlighting the interaction between its different components within the multi-task framework, is depicted in Figure \ref{fig02}. This illustration provides a clear visual representation of how the model integrates and processes various sub-tasks, contributing to its enhanced performance.

\subsection{Representation Module}
 \begin{figure}[t]
    \centering
    \includegraphics[width=\textwidth]{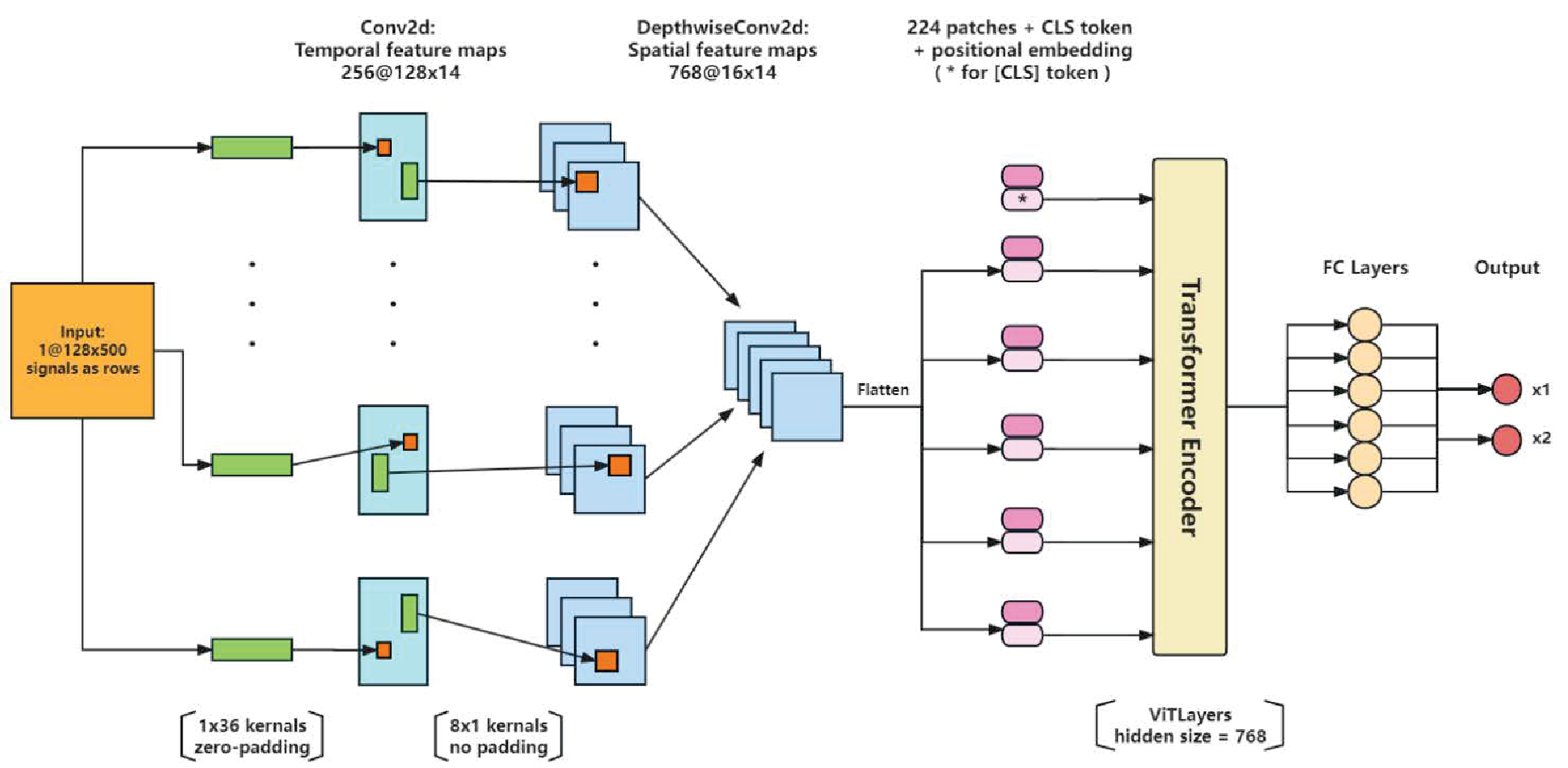}
    \caption{Model architecture of EEG2VIT \cite{yang2023vit2eeg}. In this paper, we use the Convolution Layer and Transformer Encoder as the  Representation Module }
\label{fig06}
\end{figure}

The ViT2EEG model, proposed by Yang and Modesitt \cite{yang2023vit2eeg}, utilizes a hybrid Vision Transformer architecture pre-trained on ImageNet for EEG data regression tasks. It outperforms other models, including a non-trained ViT, demonstrating that models pre-trained on image data can be effectively fine-tuned for EEG tasks.
In our model architecture, we employ the Convolution Layer and pre-trained Vision Transformer (ViT) encoder, the same as the ViT2EEG model in figure \ref{fig06}. This setup has been shown to effectively capture complex patterns in data, which is essential for the reconstruction sub-module.

The input EEG data, with dimensions \(1 \times 128 \times 500\), undergo a convolutional process yielding temporal feature maps of \(256 \times 128 \times 14\). This is followed by depthwise convolutional layers, which further refine the spatial characteristics into feature maps sized \(768 \times 16 \times 14\). The Conv2d layers utilize \(1 \times 36\) kernels with zero padding, while the DepthwiseConv2d layers apply \(8 \times 1\) kernels without padding, ensuring an effective spatial-temporal feature representation.

These features are then transformed into a sequence of \(224\) flattened patches, each integrated with a unique positional embedding. An additional [CLS] token embedding is also included, a common practice in ViT architectures to facilitate classification. The resulting embeddings are processed through a Transformer encoder, equipped with a hidden size of \(768\) to capture complex dependencies within the data.

The architecture concludes with fully connected layers, outputting two distinct values, which are the final inference results of the model. This innovative design leverages the strengths of both convolutional operations and transformer-based modeling to handle the intricacies of EEG signal analysis effectively.

\subsection{Prediction Module}

The prediction module in our architecture is designed as a sequence of interconnected layers. It comprises a fully connected layer, followed by a dropout layer for regularization, and concludes with another fully connected layer. The output of this module is articulated as follows:

\begin{equation}
\hat{y} = FC({dropout}(FC(H^{(l)}))
\end{equation}
In this formulation, $H^{(l)}$ represents the output of the last layer in the encoder. The notation $H^{(l)}$ signifies the hidden representation obtained after the input data has undergone a series of transformations through the layers of the encoder neural network. Each layer in the encoder, denoted by $l$, contributes to shaping this representation, and $H^{(l)}$ captures the information learned up to that point.
\( FC \) denotes the fully connected layers. The \( dropout \) function represents the dropout layer, a crucial component for preventing overfitting by randomly dropping units from the neural network during training.

For the main task of our model, the Mean Squared Error (MSE) loss is employed. This loss function is defined as:

\begin{equation}
Loss_{MSE}(\hat{y}, y_1) = \frac{1}{n} \sum_{i=1}^{n} (\hat{y}_i - y_{1i})^2
\end{equation}

Here, \( \hat{y} \) is the predicted output of the network, and \( y_1 \) is the actual label for the main task. The MSE loss function computes the average of the squares of the differences between the predicted and actual values, providing a measure of the model's accuracy.

This structure ensures a streamlined flow of data through the layers, facilitating effective feature extraction and subsequent prediction.

\subsection{Reconstruction Module}

The reconstruction module plays a pivotal role in our system, consisting of a series of spatial and temporal deconvolution blocks designed to incrementally expand the dimensionality of shared features to reconstruct the input data effectively.

The spatial deconvolution block is crucial for spatial feature reconstruction and is defined by the following equation:

\begin{equation}
H_{{decoder\_spatial}} = {Deconv\_spatial}(H^{(l)})
\end{equation}

In this block, \( {Deconv\_spatial} \) is composed of a three-layer structure: starting with a 1D Deconvolution layer with a kernel size of 1 \( \times \) 36. This specific configuration mirrors the first convolution layer in the encoder, ensuring symmetry in feature extraction and reconstruction. It is followed by an InstanceNorm layer, enhancing the normalization of features, and a ReLU activation layer, introducing non-linearity for better feature representation.

Similarly, the temporal deconvolution block, essential for time-series data reconstruction, is formulated as:

\begin{equation}
H_{{decoder\_temporal}} = {Deconv\_temporal}(H_{{decoder\_spatial}})
\end{equation}

The \( {Deconv\_temporal} \) block also includes three layers. It begins with a 1D Deconvolution layer, this time with a kernel size of 8 \( \times \) 1. This dimensionality aligns with the patch size used in the Vision Transformer (ViT) encoder block, allowing for a consistent approach to handling spatial-temporal data. This layer is followed by an InstanceNorm layer and a ReLU activation layer, similar to the spatial deconvolution block.

The final step in the reconstruction process is the upsampling block, defined as:

\begin{equation}
\hat{x} = {Upsampling}(H_{decoder\_temporal})
\end{equation}

This block efficiently transforms the decoder output to match the original input size, ensuring the reconstructed data \( \hat{x} \) is comparable to the original input \( x \).

Lastly, we define our loss function using Mean Squared Error (MSE) to quantify the reconstruction accuracy:

\begin{equation}
{loss\_MSE}(\hat{x}, x) = \frac{1}{N} \sum_{i=1}^{N} (\hat{x}^{(i)} - x^{(i)})^2
\end{equation}

Where \( \hat{x} \) is the reconstructed input, \( x \) is the original input, and \( N \) represents the total number of elements in \( x \). MSE is chosen for its effectiveness in emphasizing larger errors and its suitability in scenarios where maintaining the fidelity of the reconstructed data is crucial.

\subsection{Multi-Task Learning Framework}

In our multi-task learning framework, we aim to enhance the training of the primary eye-tracking task by integrating the losses from sub-tasks. The overall loss \(L\) of the framework is computed using the following equation:

\begin{equation}
L(\theta) = {Loss_{MSE}(x, y_1)} + \sum{\alpha_i Loss(x, y_i)}+\lambda||\theta||^2
\end{equation}

Where \(x\) is the input EEG signal, represented as a 2D matrix. \(y_1\) is the label of the eye-tracking task, which is also the main task. \(Loss_{MSE}(x, y_1)\) denotes the MSE loss for the eye-tracking task, and \(Loss(x, y_i)\) is the loss of other sub-task. Hyper-parameter \(\alpha_i\) is utilized to balance the relative importance of the supervised and unsupervised loss. We apply \(l_2\) regularization term with coefficient to alleviate overfitting. Our task is to minimize (\(\theta)\). All trainable parameters of the network are trained in an end-to-end manner.

To evaluate different strategies within our proposed multi-task learning framework, we developed two distinct model architectures, each focusing on a separate sub-task. The first model, named MTL-Transformer, employs a reconstruction sub-task, as introduced earlier in the paper. This model aims to reconstruct the original EEG data. The second model, MTL-Transformer2, diverges by replacing the reconstruction module with a pupil size prediction module. This auxiliary subtask was introduced to explore the relevance of pupil size to the eye-tracking task. To accommodate this, we reorganized our dataset to include pupil size for each sample. Both models were measured using the Root Mean Square Error (RMSE) metric to ensure a consistent and objective evaluation of their performance.

\section{Experiments}

\subsection{Dataset}
 \begin{figure}[t]
    \centering
    \includegraphics[width=\textwidth]{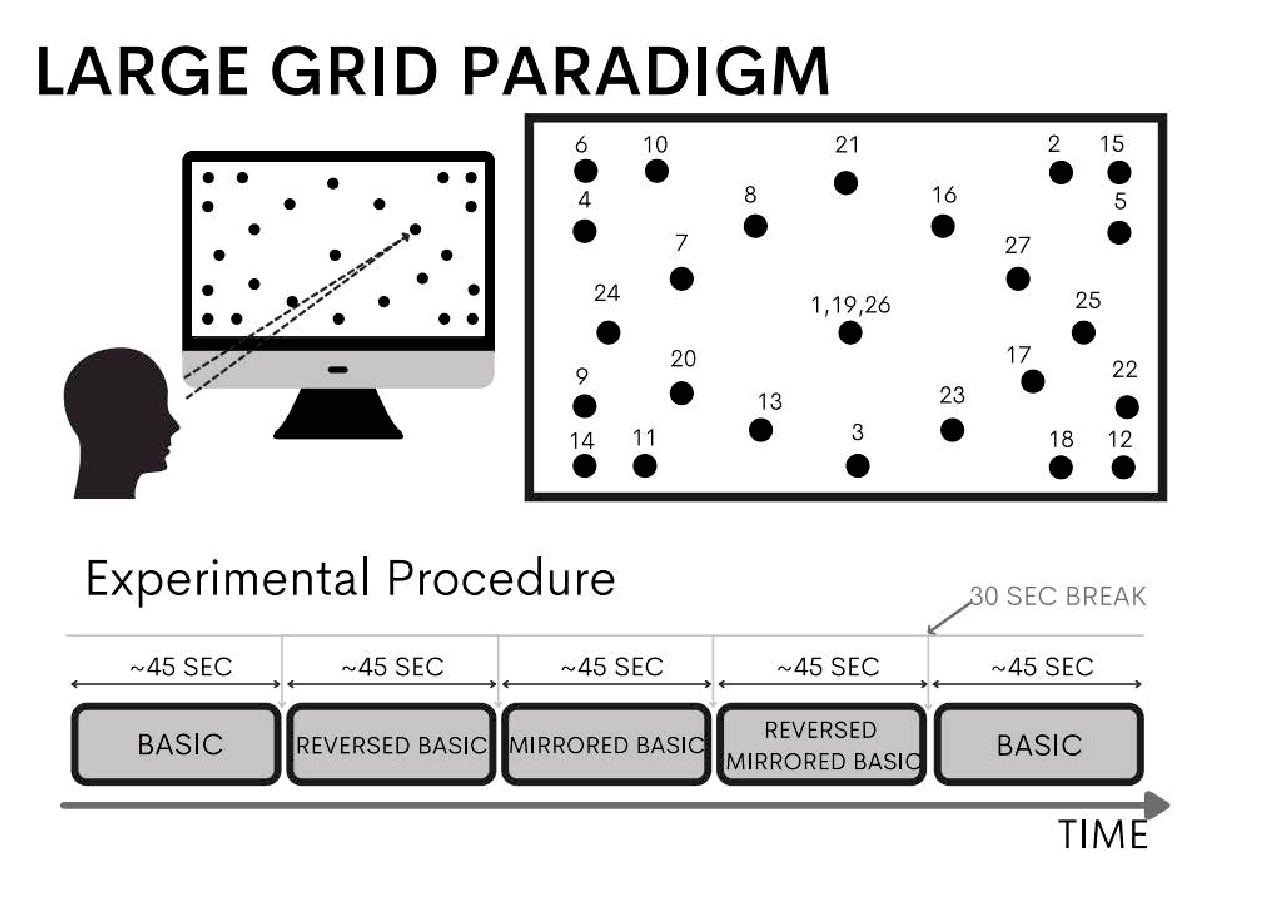}
    \caption{EEGEyeNet Large Grid Paradigm \cite{kastrati2021eegeyenet}. Participants are asked to fixate on particular dots in a given period}
\label{figure:LargeGrid}
\end{figure}

The EEGEyeNet dataset \cite{kastrati2021eegeyenet} offers a comprehensive resource for EEG research, featuring 47 hours of high-density 128-channel EEG data, which provides detailed neural activity recordings synchronized with eye-tracking data from 356 adults. This dataset is particularly suited for our study, which aims to leverage the rich EEG data to predict behavioral responses in eye-tracking tasks. Our focus on the eye-tracking task stems from its potential to reveal how neural patterns correlate with visual attention and eye movement behaviors. Detailed information about the dataset, including the specificities of the eye-tracking tasks and participant demographics, is elaborately presented in Table \ref{table01} and Figure \ref{figure:LargeGrid}. For our experiment, we propose a dataset split of 70\% for the training set, 15\% for the validation set, and 15\% for the test set. This distribution is designed to maximize learning from the EEG data while ensuring robust validation and testing of our predictive models.

\begin{table}[t]
\centering
\begin{tabular}{p{1.8cm}p{1.6cm}p{1.6cm}p{1.6cm}}
\hline
Paradigm & \# Fixations & \# Saccades & \# Blinks \\
\hline
Pro-Antisac. & 357115 & 358384 & 56179 \\
Large Grid & 68075 & 68245 & 11108 \\
VSS  & 43384 & 43443 & 971 \\
Total & 468574 & 470072 & 68258 \\
\hline
\end{tabular}
\caption{Eyes event label distribution in EEGEyeNet dataset (minimal preprocessing) \cite{kastrati2021eegeyenet}}
\label{table01}
\end{table}

\subsection{Baseline Models}

\subsubsection{Machine Learning}
We employed a range of traditional machine-learning algorithms as baseline models. These include K-Nearest Neighbors (KNN), Support Vector Machines with Radial Basis Function (RBF SVC/SVR), Linear Regression, Ridge Regression, Lasso Regression, Elastic Net, Random Forest, Gradient Boosting, AdaBoost, and XGBoost. While these methods provide solid benchmarks, they have limitations in handling the high dimensionality and complex temporal dynamics inherent in EEG data.

\subsubsection{Deep Learning}
\textbf{Convolutional Neural Network (CNN)}
CNNs were effective in capturing spatial patterns in EEG data but less adept at modeling temporal dynamics.

\textbf{PyramidalCNN}
The PyramidalCNN, with its unique structure, offered improved performance in capturing hierarchical features, leading to better generalization \cite{johnson2017deep}.

\textbf{EEGNet}
EEGNet, designed for EEG data analysis, showed proficiency in handling both spatial and temporal features but may struggle with very large datasets \cite{lawhern2018eegnet}.

\textbf{InceptionTime}
InceptionTime's modular architecture allowed for a robust capture of temporal dynamics, surpassing traditional CNNs \cite{ismail2020inceptiontime}.

\textbf{Xception}
Xception's depthwise separable convolutions were efficient, though they may not fully exploit the multi-channel nature of EEG data \cite{chollet2017xception}.

\textbf{EEGViT}
The EEGViT, adapting the Vision Transformer for EEG data, presented an innovative approach in modeling long-range dependencies, a common challenge in EEG analysis \cite{yang2023vit2eeg}.

\subsection{Implementation Details}
All models in our study were trained for 15 epochs on an RTX 4090 GPU. For deep learning models, we set an initial learning rate of \(10^{-4}\) and implemented a decay strategy, reducing the learning rate by 10\% every 6 epochs. This approach is designed to balance the rate of convergence and ensure effective learning over the training period.

In our proposed model, we integrated two dropout layers with a dropout rate of 0.3, specifically in the prediction module. This rate is higher than typical settings, chosen to mitigate overfitting while dealing with the complex nature of EEG data. This dropout strategy is particularly crucial given the model's architecture and the high-dimensional feature space of the EEG signals.

\section{Results}

\begin{table}[t]
\centering
\begin{tabular}{|l|c|}
\hline
\textbf{Model} & \textbf{\shortstack{AbsolutePosition\\RMSE (mm)}} \\
\hline
Naive Baseline & 123.3 $\pm$ 0 \\
\hline
KNN & 119.7 $\pm$ 0 \\
RBF SVC/SVR & 123 $\pm$ 0 \\
Linear Regression & 118.3 $\pm$ 0 \\
Ridge Regression & 118.2 $\pm$ 0 \\
Lasso Regression & 118 $\pm$ 0 \\
Elastic Net & 118.1 $\pm$ 0 \\
\hline
Random Forest & 116.7 $\pm$ 0.1 \\
Gradient Boost & 117 $\pm$ 0.1 \\
AdaBoost & 119.4 $\pm$ 0.1 \\
XGBoost & 118 $\pm$ 0 \\
\hline
CNN & 70.2 $\pm$ 1.1 \\
PyramidalCNN & 73.6 $\pm$ 1.9 \\
EEGNet & 81.7 $\pm$ 1.0 \\
InceptionTime & 70.8 $\pm$ 0.8 \\
Xception & 78.7 $\pm$ 1.6 \\
\hline
ViT-Base & 61.5 $\pm$ 0.6 \\
ViT-Base Pre-trained & 58.1 $\pm$ 0.6 \\
EEGViT & 61.7 $\pm$ 0.6 \\
EEGViT Pre-trained & 55.4 $\pm$ 0.2 \\
\hline
\textbf{MTL-Transformer(Ours)} & \textbf{54.1 $\pm$ 0.2} \\
\textbf{MTL-Transformer2(Ours)} & \textbf{57.4 $\pm$ 0.3} \\
\hline
\end{tabular}
\caption{Root Mean Squared Error (RMSE) Comparison of Baseline Models on the EEGEyeNet eye-tracking task \cite{kastrati2021eegeyenet}. The primary model, MTL-Transformer, demonstrates significant performance improvement, utilizing EEGViT Pre-trained as its base model. Additionally, MTL-Transformer2, which includes pupil size prediction as an auxiliary sub-task, is presented to demonstrate the scope of our experimental exploration despite its lesser impact on RMSE reduction.}
\label{table02}
\end{table}

 \begin{figure}[t]
    \centering
    \includegraphics[width=\textwidth]{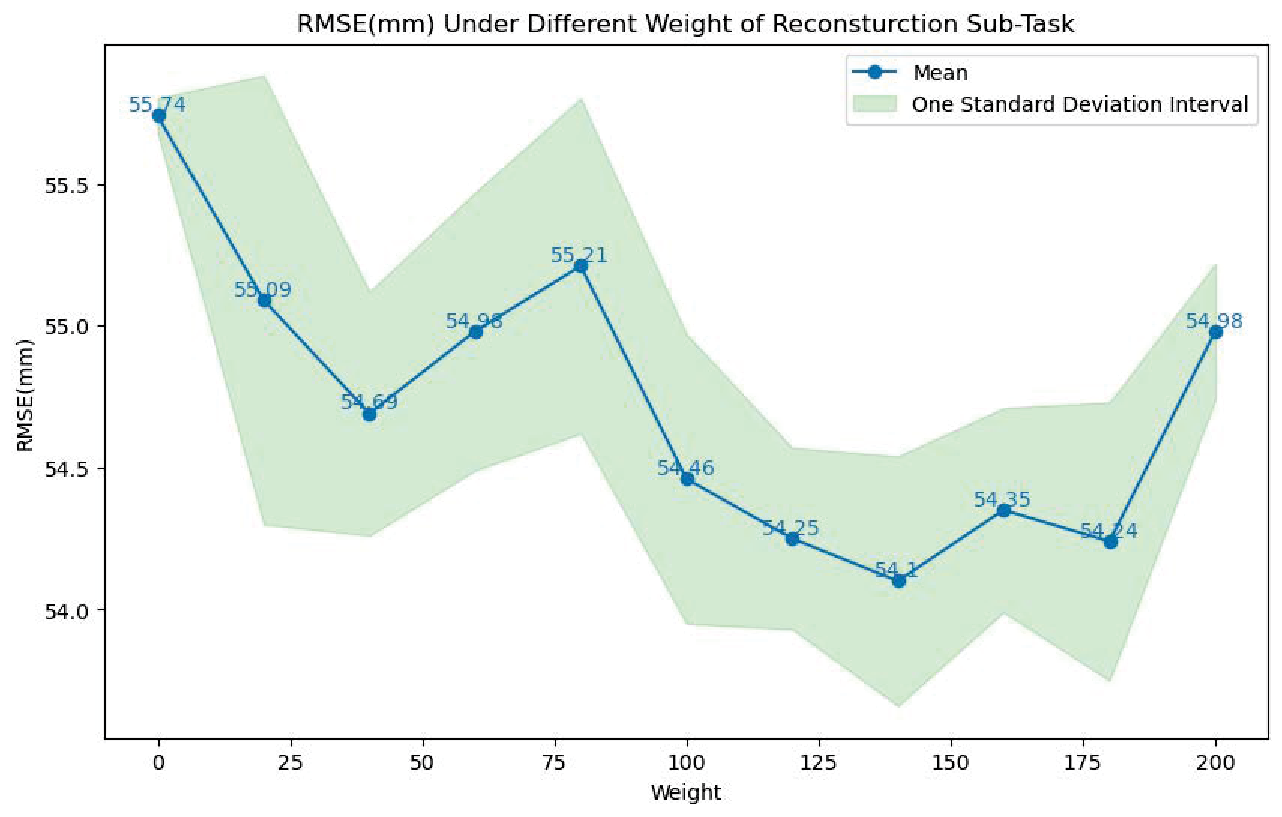}
    \caption{RMSE(mm) Under Different Weight of Reconstruction Sub-Task}
\label{fig07}
\end{figure}

 \begin{figure}[t]
    \centering
    \includegraphics[width=\textwidth]{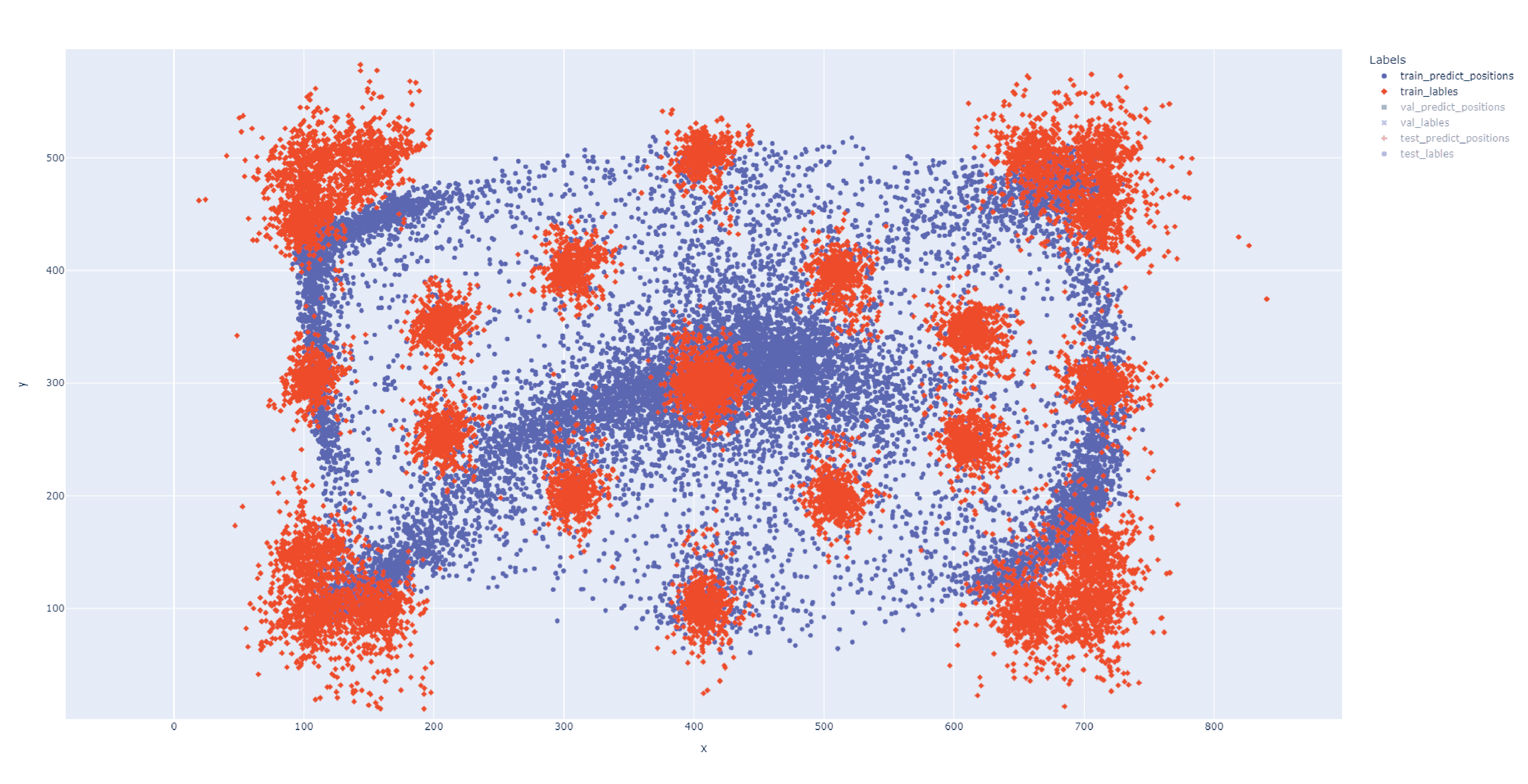}
    \caption{Predict Gazing Position and Real Gazing Position on Training Dataset}
\label{fig08}
\end{figure}

The performance of various models on the EEG eye-tracking task is summarized in Table \ref{table02}. Notably, our proposed model achieved a Root Mean Square Error (RMSE) of 54.1mm, which slightly surpasses the current State-Of-The-Art (SOTA) model's RMSE of 55.4mm. This improvement, although marginal, indicates the effectiveness of our model's architecture and the methodologies employed, especially in handling the complexities of EEG data in eye-tracking tasks.

Figure \ref{fig07} illustrates the RMSE of our model under varying weights assigned to the reconstruction sub-task. At a weight of 0, the reconstruction sub-module does not participate in the gradient computation, and our model's results align with the EEGViT Pre-trained model, which is the base model. This parallel performance indicates that the enhancements in accuracy are not attributable to alterations in the model's structure. As the weight increases to 140, there is a discernible improvement in model accuracy, suggesting that the reconstruction sub-module contributes positively to the base model's performance. However, beyond a weight of 140, the excessive emphasis on the reconstruction sub-task seems to detract from the sub-task at hand, as evidenced by a decline in accuracy. This trend demonstrates a critical balance between the reconstruction weight and the model's focus on the primary task, underscoring the need for optimal weight tuning to harness the reconstruction sub-module's benefits without compromising the main objective.

\section{Discussion}

Our research presents promising implications for the field of EEG-based eye tracking. The slight yet significant improvement in accuracy provided by our model paves the way for more precise and reliable EEG eye-tracking systems. This advancement is particularly relevant in applications where minute differences in eye movement can have substantial implications, such as in neuromarketing or neurological disorder diagnosis.

Figure \ref{fig08} shows the eye-tracking prediction on the training dataset. Our model demonstrates a commendable capacity to discriminate between central and peripheral points. However, it exhibits limitations in accurately distinguishing points within the intermediate regions, with a predominant aggregation of data at the center. This phenomenon may be attributed to a frequency bias towards central points, resulting in an imbalance within the dataset. To address this, future iterations of our model could incorporate weighted learning for different regions, facilitating a more balanced and nuanced understanding and thereby enhancing the model's predictive accuracy.

Looking ahead, we aim to broaden the applicability of our model by testing it on various other EEG datasets. Such an expansion will not only validate the model's effectiveness across different data types but also enhance its robustness and generalizability. This step is crucial for asserting the model's utility in diverse real-world scenarios.

Additionally, the versatility of our Multi-task Learning module is a notable aspect of our architecture. Its design allows it to be integrated as a separate module into any EEG-based task. This modular approach offers a flexible solution for improving existing EEG analysis systems, potentially transforming how EEG data is processed and interpreted in various applications.

Moreover, we attempted to leverage pre-trained language Transformer models, such as GPT and BERT, which are typically used for time-series or language tasks. However, these models generally demand substantial GPU memory capacity, which exceeds the capabilities of our personal workstations. This limitation constrained the scope of our experiments. Future work will, therefore, focus on optimizing computational efficiency, perhaps through model distillation or pruning techniques that can reduce the memory footprint of these large models. Future studies can also investigate other potential deep learning approaches on various datasets for comparative analysis \cite{an2023transfer,an2023survey,jiang2023successfully,gui2024remote,lu2023machine,chen2024trialbench,ma2022traffic,ma2024data,tan2023audio,tan2021multivariate,qiu2023modal,zhao2024deep,zhang2022attention,zhang2023trep}.

In conclusion, while our current focus has been on EEG eye-tracking tasks, the broader impact of our work lies in its potential to revolutionize various aspects of EEG data analysis and application. Future research will delve deeper into these possibilities, continually pushing the boundaries of what is achievable in this domain.

\section{Conclusion}

In this study, we have demonstrated the effectiveness of integrating multi-task learning with Vision Transformers in the domain of EEG eye-tracking. Our approach has successfully employed an innovative EEG signal reconstruction sub-module, enhancing the feature extraction capabilities of deep learning models applied to this task. This sub-module, adaptable to various Encoder-Classifier-based models, facilitates end-to-end training within a multi-task learning framework and operates effectively under unsupervised learning conditions.

Our experimental results, particularly the achieved RMSE of 54.1mm, which surpasses the previous state-of-the-art model, underscore the potential of our method in improving EEG-based eye-tracking systems. This advancement is not only significant in terms of model performance but also in its application potential across various EEG datasets and tasks.

Looking forward, the adaptability and versatility of our Multi-task Learning module open new avenues for enhancing EEG data processing and interpretation. This work lays the groundwork for future research in this area, aiming to further explore and expand the capabilities of deep learning models in the realm of neuroscience and cognitive research. We believe that our approach represents a novel paradigm in EEG data analysis, with the potential to contribute significantly to various EEG-related challenges and applications.

\bibliographystyle{splncs04}
\bibliography{Enhancing_Eye-Tracking_Performance_through_Multi-Task_Learning_Transformer}
%




\end{document}